\documentclass[floatfix,10pt,aps,prb,twocolumn,longbibliography,a4paper]{revtex4-2}

\usepackage[english]{babel}
\usepackage{float}
\usepackage{amsmath}
\usepackage{graphicx}
\usepackage[colorlinks=true, allcolors=blue]{hyperref}

\begin{document}
\title{Effect of biquadratic magnetic exchange interaction in the 2D antiferromagnets $M$PS$_3 (M=$Mn, Fe, Co, Ni)}
\author{M. Amirabbasi}
\author{P. Kratzer}
\email{Peter.Kratzer@uni-due.de}
\affiliation{Fakult\"at f\"ur Physik and CENIDE, Universit\"at Duisburg-Essen, Lotharstra{\ss}e 1, 47057 Duisburg, Germany}

\begin{abstract}
The two-dimensional van der Waals (vdW) materials $M$PS$_{3} \, (M=$ Mn, Fe, Co, Ni) display antiferromagnetic ordering of the magnetic moments at the transition metal ions. 
The possibility to exfoliate thin layers that preserve the magnetic order makes these materials interesting for numerous applications in devices that require integration of flexible patches of magnetic materials, e.g. in antiferromagnetic spintronics. Hence, an improved understanding of their magnetic properties is desirable.  
Here, we parameterize spin Hamiltonians for a monolayer of all four materials of this class using density functional theory plus Hubbard $U$  calculations.  
We provide a step-by-step guide for calculating the magnetic exchange interactions and magnetic anisotropy energy using the (non-)collinear DFT+\textit{U}(+ SOC) approach with a suitably chosen $U$ for each material. 
It is found that the biquadratic interactions gain in importance while moving through the 3d series. Retaining the leading terms of a Holstein-Primakoff-transformed spin Hamiltonian, the magnon spectra are calculated.  While MnPS$_3$ is found to be an almost isotropic antiferromagnet  with a tiny gap, the biquadratic interaction opens an increasingly wider gap for FePS$_3$, CoPS$_3$ and NiPS$_3$. In line with this observation, Monte Carlo simulations demonstrate that the biquadratic interactions contribute to a systematic rise in the N\'eel temperature from FePS$_3$ to NiPS$_3$.  
\end{abstract}

\maketitle
\section{Introduction}

Transition metal phosphorus trisulfides, denoted as $M$PS$_3$ where M can be Mn, Ni, Fe, or Co, represent a class of materials that have gained significant attention \cite{Du2016,Mayorga2017,Wang2018AdvFuncMat} for their remarkable electronic, magnetic, and optical characteristics. These materials are particularly noted for their impressive optoelectronic properties \cite{Yang2022}, including strong absorption in the visible to near-infrared spectrum and efficient charge separation and transport\cite{Musari_2022}. They feature a layered structure similar to graphene, which not only piques interest for fundamental research but also facilitates their exfoliation into two-dimensional layers. These thin layers often exhibit properties distinct from their bulk counterparts, offering a rich avenue for investigating new physical phenomena and potential technological applications.

Magnetism is a subtopic in the field of 2D materials that has drawn significant interest recently \cite{acsnano.genome}.
The magnetic nature of $M$PS$_3$ is a key aspect that this article aims to explore in depth. By calculating the spin Hamiltonian and analyzing the magnon spectrum, we seek to unravel the magnetic properties of these materials. The study of the magnon spectrum in $M$PS$_3$ is not just a topic of fundamental interest; it also has significant implications for the fields of spintronics\cite{Ahn2020} and magnonics\cite{Bostrom2023PRL}, where electron spins and magnons are utilized for advanced information processing and storage. The magnon spectrum plays a vital role in determining the magnetic behavior of a material, such as its magnetic ordering temperature and characteristics crucial for spintronic applications. The spin Hamiltonian, which includes both isotropic and anisotropic exchange interactions, provides insights into the interactions between magnetic moments and their tendencies to align in specific directions.
The unique layered structure and magnetic properties of $M$PS$_3$ materials offer a template for designing new materials with tailored properties. By understanding the interplay between structure, magnetism, and electronic properties in these materials, researchers can engineer new compounds with desired functionalities for specific applications.

In the realm of 2D monolayer systems governed by a Heisenberg Hamiltonian, which inherently encompasses short-range magnetic interactions and preserves spin rotational symmetry, a theorem proposed by Mermin and Wagner \cite{Mermin1966}  precludes the establishment of long-range ferromagnetic (FM) or antiferromagnetic (AFM) ordering at any finite temperature. This theorem's foundation rests on the inherent characteristics of the isotropic Heisenberg Hamiltonian, notably its continuous symmetry, which facilitates the existence of long-wavelength spin waves without an energy gap. These spin waves, owing to their gapless nature, are thermally excitable at any finite temperature, posing a significant challenge to the sustenance of long-range magnetic order in low-dimensional structures.
In contrast, scenarios that involve a breach in the spin rotational invariance, exemplified by anisotropic magnetic interactions within the framework of a two-dimensional Ising model, alter this paradigm. In such cases, the introduction of anisotropy leads to the formation of an energy gap in the spin wave spectrum. This gapped spin wave spectrum plays a crucial role in stabilizing long-range magnetic order by diminishing the influence of thermal fluctuations. Such stabilization becomes prominent below a specific transition temperature, marking a stark divergence from the behavior predicted by the isotropic Heisenberg model in two-dimensional systems. A principal objective of the research outlined in this document is to examine the presence and implications of anisotropic exchange interactions within $M$PS$_{3}$ materials, with specific focus on phenomena such as single-ion anisotropy (SIA) and Dzyaloshinskii-Moriya interactions (DMI)\cite{Moriya1960}. This investigation is crucial in resolving the key question of whether two-dimensional (2D) $M$PS$_{3}$ materials can retain their magnetic order during the transition from bulk to monolayer structures.
Additionally, in this study, we are examining the impact of biquadratic interactions in $M$PS$_{3}$ materials. Since these materials have a collinear magnetic order, we expect that the biquadratic interaction parameter may have a negative value. This prediction aligns with the theoretical framework outlined in Eq.~(\ref{H}) of the paper. Understanding these biquadratic interactions is crucial for gaining a deeper understanding of the complex magnetic properties of these materials, particularly as they are reduced to lower-dimensional states.

\begin{figure}[!htp]
    \centering
   \includegraphics[width=0.65\columnwidth]{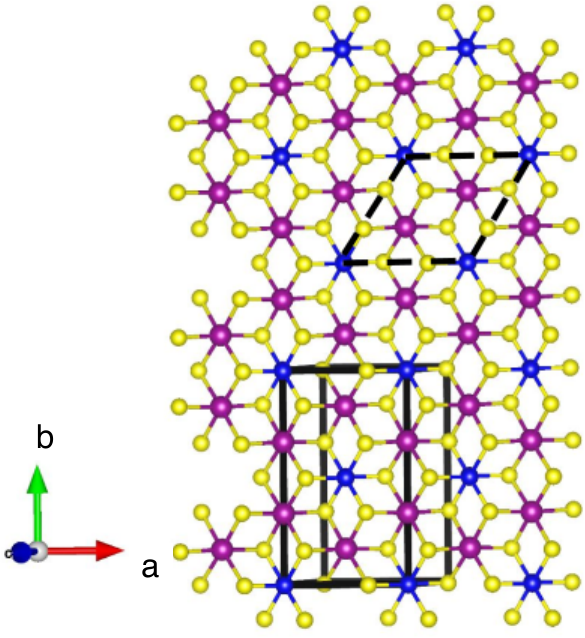}
\caption{\label{fig:monoclinic} The experimental CIF file of \( M \)PS\(_{3} \) monolayer designates the solid rectangle spanned by $a$ and $b$ as the simplest unit cell for the \( C2/m \) space group. The dashed lines outline the rhombus-shaped primitive cell, spanned by $a$ and $b'$ which comprises 10 ions. 
To maintain uniformity across calculations, other supercells are generated based on this primitive cell.
} 
\end{figure}

The compound \(\mathrm{MPS}_3\) crystallizes in a monoclinic structure with the \(C2/m\) space group symmetry. The MAGNDATA webpage \cite{magndata1} provides experimental CIF files of \(\mathrm{MPS}_3\) materials. These files represent the magnetic unit cell, which includes 20 ions (four \(M\)-ions, four phosphorus ions, and twelve sulfur ions).
The material's bulk structure is made up of layers stacked together and held in place by van der Waals forces. Within each layer, a transition metal ion $M$ is surrounded by six sulfur atoms, forming a distorted octahedral structure that contributes to the formation of a hexagonal lattice interconnected by S ions. Central to these hexagonal arrangements are two phosphorus ions, each bonded to three sulfur ions.
The basal plane of the monoclinic cell is spanned by two orthogonal lattice vectors $a$ and $b$ (cf. Fig.~\ref{fig:monoclinic}). Since we are interested in 2D layers, the length and angle of the third lattice vector $c$ doesn't matter for our calculations. For the electronic structure calculations, we first define 
a primitive cell, which includes only two \(M\)-ions, two phosphorus ions, and six sulfur ions. It is spanned by two lattice vectors $a$ and $b'$ forming an angle of 60$^\circ$, with $|b'| \approx |b|/\sqrt{3}$. 
 This configuration maintains all point symmetry operations of the \(C2/m\) space group. In the monoclinic system, the presence of a two-fold rotation axis and a mirror plane are essential symmetry elements. 
 Our primitive cell ensures that these symmetries are preserved, reflecting the structural characteristics of the \(C2/m\) space group.
 In a later step, supercells are built out of this primitive cell as required to model certain spin arrangements.
The magnetic moments in $M$PS$_3$, originating from the unpaired d-electrons of the M ions, interact with each other, leading to the emergence of magnetic order. This aspect of magnetic interaction and order is a cornerstone of the intriguing properties exhibited by $M$PS$_3$ materials, underlining their potential for a wide array of applications in the field of materials science.
%#############################################
%################################################

%################################################
\begin{figure}[!htp]
    \centering
   \includegraphics[width=\columnwidth]{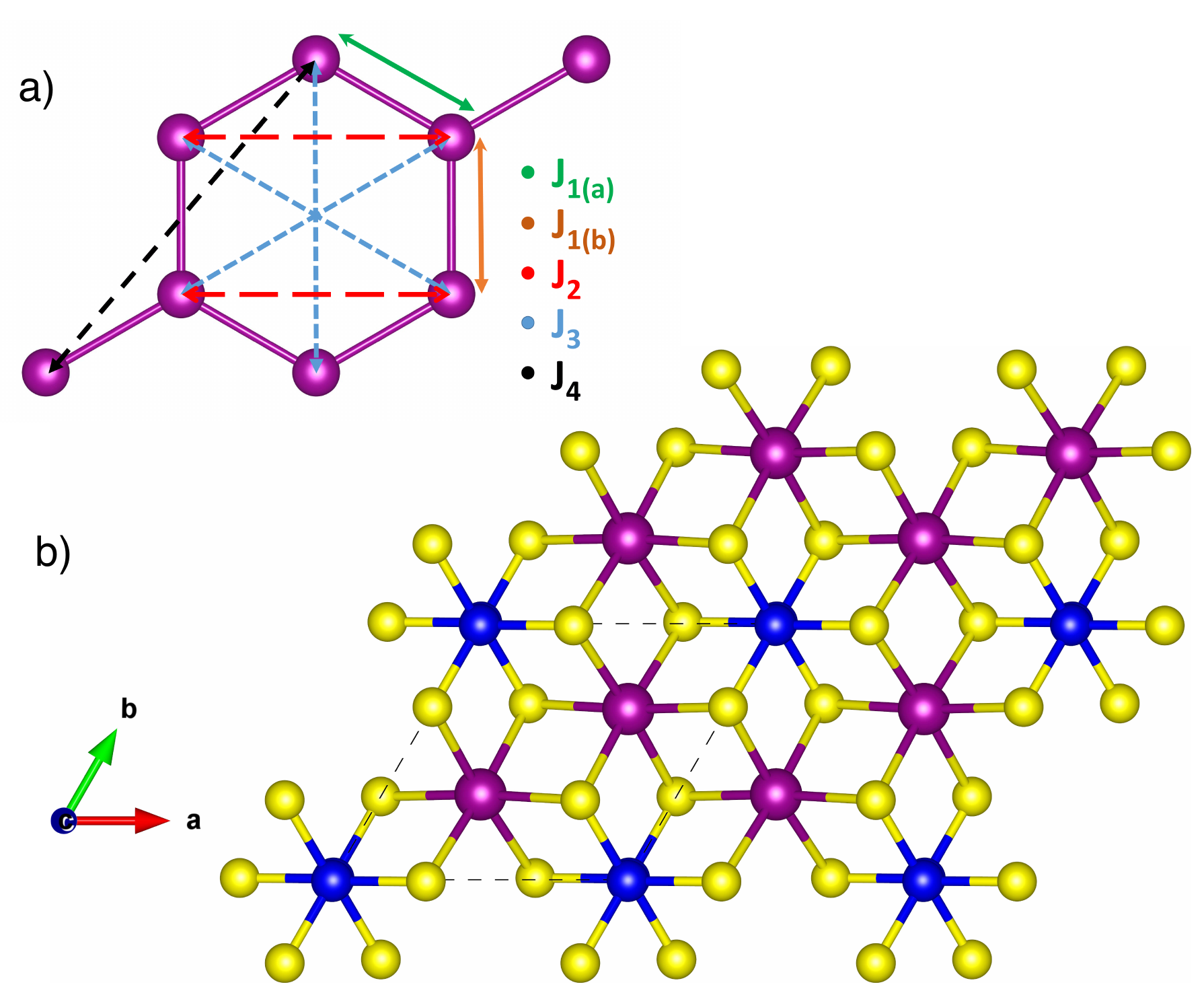}
\caption{\label{fig:geometry} (Color online)  (a) Schematics of exchange interactions for different neighbors. (b) $2 \times 2$ supercell of a $M$PS$_{3}$ monolayer. The purple, blue, and yellow spheres are transition metals, P and S ions, respectively. The dashed line shows the primitive cell containing two $M$ ions. }
\end{figure}
%%%%%%%%%%%###########################################
\begin{figure*}[!htp]
    \centering
      \includegraphics[width=\textwidth]{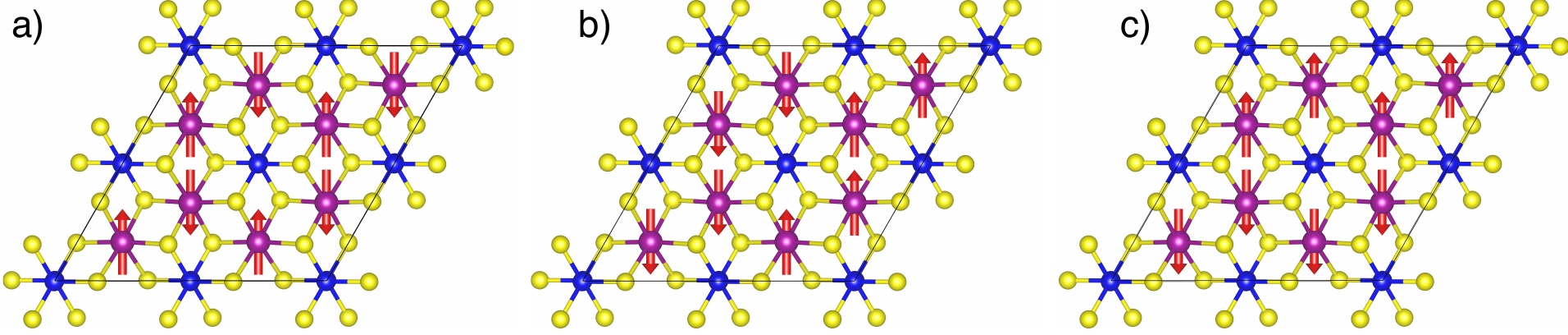}
\caption{\label{fig:spin_pattern} (Color online)  Different AFM ground states: (a) N\'eel ground state of MnPS$_{3}$ (b)~Long-bond zigzag found in FePS$_{3}$  (the vertical Fe -- Fe spacing in the figure is the 'long' one), and (c)~Short-bond zigzag found in CoPS$_{3}$  and NiPS$_{3}$. It should be noted that according to our GGA+$U$+SOC calculations, 
the $a-b$ plane is the easy plane for Mn (total energy when the spins align in $b$-axis is lower, but compared to $a$-axis, the total energy difference is very small. Hence, one can say $a-b$ plane is an easy plane). For Fe, the spins prefer to align along the $c$-axis which is perpendicular to the $a-b$ plane. For Co and Ni, $b$-axis and $a$-axis are easy axes, respectively.
}
\end{figure*}
%%%%%%%%%%%%%%%%%%%%###########################################3
The paper is structured as follows: In Section~\ref{sec:methods} we provide details of the electronic structure calculations and Monte Carlo (MC) simulations.
In Section~\ref{sec:results} we first report on the structural optimization and magnetic order.
We then describe the electronic structure before detailing the magnetic interactions.
Finally, in Section~\ref{sec:conclusion} we discuss the implications of our results for $M$PS$_{3}$ single-layers and in the broader context of magnetic vdW layered materials.

\section{Methods}
\label{sec:methods}
The present study employs Density Functional Theory (DFT) to investigate the magnetic and structural properties of materials. DFT is a well-established computational approach for the evaluation of electronic properties of materials. However, it may not accurately capture the electronic structure of materials with strongly correlated electrons, such as transition metal oxides or rare earth magnets, using standard DFT. To address this issue, the DFT+$U$ method is commonly utilized, which incorporates an on-site Coulomb interaction term ($U$) to better account for the electronic interactions within the material.
To apply the DFT+$U$ method to investigate magnetic properties, the appropriate value of the $U$ parameter must first be determined. This value can be derived experimentally or from previous theoretical studies. It is crucial to choose the correct $U$ value, as it directly affects the computed magnetic properties of the material.
In this work, the DFT calculations are performed using the Quantum Espresso\cite{Giannozzi_2009} (QE) and the all-electron FLEUR\cite{fleur, wortmann2023fleur} code. In the QE calculations, the wavefunctions and charge density are expanded in plane waves using a cutoff of 50 Ry and 550 Ry, respectively.
The present study utilizes the FLEUR-based calculations to investigate the noncollinear($+$SOC) magnetic properties of $M$PS$_{3}$. The wave function expansion cutoff in the interstitial region is set to $k_\text{max} =$ 3.8 a.u.$^{-1}$

%%%%%%%%%%%%%%%%%%%%%%%%%
\begin{table}[!htp]
  \centering
  \caption{Lattice constants of optimized geometry in GGA+$U$ 
  for the crystallographically primitive cell spanned by $a$ and $b'$. Note that the conventional monoclinic lattice vector $|b| = \sqrt{4 b'^2 - a^2}$.} The optimum $U$ values have been chosen through comparing the calculated bandgap with its experimental values.
  \begin{tabular}{ccccc}
    \hline
    Material & $U_{\text{eff}} \, (\text{eV})$ & $a \, (\text{\AA})$ & $b' \, (\text{\AA})$ & $E_{\mathrm{g}} \, (\text{eV})$ \\
    \hline
    MnPS$_{3}$& 3.0 & 6.193 & 6.193 &  2.0  \\
    MnPS$_{3}$& \text{exp.} & 6.076 & 6.076 &  2.94 \cite{Grasso1991,Du2016} \\
    \hline
    FePS$_{3}$& 2.22 & 6.017 & 6.052 &1.23 \\
    FePS$_{3}$& \text{exp.} & 5.940 & 5.972 & 1.23\cite{Ramos2021}, 1.44\cite{Budniak2022} \\
    \hline
    CoPS$_{3}$& 3.0 & 5.954  & 5.954 &  1.35  \\
    CoPS$_{3}$& \text{exp.} & 5.901 & 5.901 &  1.5\cite{Jin2022}  \\
    \hline  
    NiPS$_{3}$& 5.7 & 5.828 &  5.829&  1.89 \\
    NiPS$_{3}$& \text{exp.} & 5.812 &5.813 &   1.6\cite{Du2016} \\
    \hline    
  \end{tabular}
  \label{tab1}
\end{table}

%%%%%%%%%%%%%%%%%%%%%%%%%%%%%%%%%%%%%%%%%%%%%%%%%%%%%%%%%%%%%%%%%%%%%%%%%%%%%%%%%%%%%%%%%%%%%%%%%%%%%%%%%%
Semicore states, specifically the 3s and 3p orbitals of the transition metal, are included in the calculations. In the selection of Muffin-tin radii for magnetic ions across various systems, considerations are made regarding the influence of semicore electrons and lattice constants. As delineated in Tab.~\ref{tab1}, the lattice constants exhibit a decremental trend from Mn to Ni. In alignment with this observation, the Muffin-tin radii for Mn, Fe, Co, and Ni have been determined as 2.9, 2.8, 2.7, and 2.6 atomic a.u., respectively. To enhance the precision in the comparative analysis of calculations, the Muffin-tin radii for P and S have been consistently set at 1.90 a.u. and 1.49 a.u., respectively, across all studied materials.  The spin magnetic moments from DFT$+U$ calculations following Dudarev's approach \cite{Dudarev1998PhysRevB} by choosing these values are found to be in reasonable agreement with the values predicted by Hund's rule. The exchange-correlation energy is approximated using the generalized gradient approximation (GGA) in the Perdew-Burke-Ernzerhof parameterization PBE~\cite{Perdew1996}.

To determine the type of magnetic order of the ground state, we define a model spin Hamiltonian :
\begin{align}
H_{\rm {spin}} & = H_{\rm Heis} %-\frac{1}{2}\sum_{i\neq j} J_{ij}(\vec{S_{i}}\cdot\vec{S_{j}})
    +\frac{1}{2} B\sum_{\rm n.n} (\vec{S_{i}}\cdot\vec{S_{j}})^{2} \nonumber \\
  & +\frac{1}{2} D \sum_{\rm n.n} \hat{D}_{ij}\cdot(\vec{S_{i}}\times \vec{S_{j}})+\Delta\sum_{i} (\vec{S_{i}}\cdot\vec{d_{i}})^{2}
\label{H}
\end{align}
where $\vec{S_\text{i}}$ represents the direction of magnetic spins, $H_{\rm Heis}$ is the usual Heisenberg Hamiltonian, %(for details see below), 
$B$, $D$ and $\Delta$ are the strengths of bi-quadratic, DMI and SIA, respectively.
Moreover, unit vectors $\hat{D}_\text{ij}$ and $\vec{d_\text{i}}$ show the direction of the DMI  and the easy axis of magnetization at each site $i$, respectively.
It should be noted that the direction of DMI is determined by Moriya rules~\cite{Moriya1960}. Due to the centrosymmetric 2/m point group symmetry,  the $M$PS$_{3}$ monolayer has a mirror plane perpendicular to the $b$-axis.
According to the Moriya rules, when a mirror plane includes two ions, the $D$ vector should be perpendicular to the mirror plane. Magnetic interactions were obtained by fitting a model Hamiltonian to total energy calculations for various magnetic configurations, as described in the appendix.
The Heisenberg Hamiltonian is given by
\begin{align}
H_{\rm Heis} & =  -\frac{1}{2}\sum_{i\neq j} J_{ij}(\vec{S_{i}}\cdot\vec{S_{j}})
\end{align}
Motivated by our previous work~\cite{Amirabbasi2023} 
on orbital ordering in FePS$_3$, we distinguish between 'close' first neighbors with exchange parameter $J_{1a}$ and 'long-bond' first neighbors with $J_{1b}$ if the distances between the $M$ atoms differ by more than 0.05 \AA.  In addition, we include interactions up to the forth-nearest neighbors to ascertain the convergence of the expansion. 
Consequently, the calculations for determining the $J$ parameters require a 2$\times$2 cell (with 40 atoms), and we use a 10$\times$10$\times$1 Monkhorst-Pack $k$-point mesh\cite{MoPa76}.
The other exchange parameters $B$, $D$ and $\Delta$ were obtained from a primitive cell (with 10 atoms), and we use a 20$\times$20$\times$1 optimized Monkhorst-Pack $k$-mesh.

We conduct Monte Carlo simulations with the Esfahan Spin Simulation package (ESpinS)\cite{Rezaei2019ESpinSAP}, treating the spin as a classical vector of unit length, using the replica exchange method on a 30$\times$30 simulation cell containing 3600 spins. Each spin is subjected to 2$\times$10 steps at each temperature. To minimize the correlation between successive data, we collect statistics every 10 MC steps. The crystal structure figures are generated using VESTA software~\cite{vesta}.

Magnon spectra are calculated analytically by re-writing the spin operators by bosonic operators using the Holstein-Primakoff transformation\cite{HoPr40}. More details are given in the appendix and in Ref.s~\onlinecite{Lee2018,Bezazzadeh2021,Olsen}. 

\section{Results and discussion}
\label{sec:results}
In this section, we present the results of our study and provide a comprehensive discussion of their implications for the field.
\subsection{Electronic structure}
In our study, we observe the progressive filling of the d-shell across MnPS$_3$, FePS$_3$, CoPS$_3$, and NiPS$_3$. In the context of our study, the spin and orbital magnetic moments     
  (Tab.~\ref{tab3}) are pivotal in understanding their magnetic properties. For MnPS$_3$, the Mn\(^{2+}\) ion with a 3d\(^5\) configuration exhibits a high spin state, in line with Hund's rule, leading to a spin moment of 5 $\mu_\mathrm{B}$ and a negligible orbital moment. Moving to FePS$_3$, the Fe\(^{2+}\) ions with a 3d\(^6\) configuration show a spin moment of 4 $\mu_\mathrm{B}$, consistent with Hund's rule, and the largest orbital moment. In CoPS$_3$, Co\(^{2+}\) ions feature a 3d\(^7\) configuration, resulting in a 3 $\mu_\mathrm{B}$ spin moment. Finally, NiPS$_3$, with Ni\(^{2+}\) ions having a 3d\(^8\) configuration, shows a reduced spin moment of 2 $\mu_\mathrm{B}$. These theoretical predictions, based on the oxidation state and electron configurations, provide a framework for understanding the magnetic behavior of these compounds, although experimental validation is crucial for a comprehensive understanding.

%%%%%%%%%%%%%%%%%%%%%%%%%%%%%%%%%%%%%%%%%%%%%%%%%%%%%%%%%%%%%%%%%%%%%%%%%%%%%%%%%%%%%%%%%%%%%%%%%%%%%%%%
\begin{table}[htb]
    \centering
        \caption{Spin and orbital moments of $M$PS$_{3}$ 2D magnets.}
    \begin{tabular}{ccccc}
        \hline
         Material& Spin moment ($\mu_\mathrm{B}$) & Orbital moment($\mu_\mathrm{B}$)\\
             \hline
         MnPS$_{3}$(3d$^{5}$)&4.68 & 0.02\\
         \hline
         FePS$_{3}$(3d$^{6}$)&3.61&0.77\\
         \hline
         CoPS$_{3}$(3d$^{7}$)&2.51&0.22\\
         \hline
         NiPS$_{3}$(3d$^{8}$)&1.52&0.11\\
         \hline
    \end{tabular}
    \label{tab3}
\end{table}
%%%%%%%%%%%%%%%

In this paper, we observe distinct structural and magnetic properties across MnPS\(_{3}\), FePS\(_{3}\), CoPS\(_{3}\), and NiPS\(_{3}\). These materials adhere to an ideal honeycomb lattice structure, whereas FePS\(_{3}\) exhibits a notable deviation with its distorted honeycomb lattice. This unique distortion~\cite{Lancon2016} in FePS\(_{3}\) at low temperatures leads to a significant difference between the long-bond and short-bond in its zigzag ground state.\cite{Amirabbasi2023} 
Remarkably, MnPS\(_{3}\) exhibits a N\'eel AFM ground state, contrasting with NiPS\(_{3}\) and CoPS\(_{3}\), which both display a zigzag AFM ground state. For FePS\(_{3}\), the magnetic order at low temperatures is characterized as a long-bond zigzag, highlighting the impact of lattice distortion on its magnetic properties. These findings elucidate the complex relationship between lattice structure and magnetic behavior in transition metal phosphosulfides. The different magnetic orders are shown in Fig.~\ref{fig:spin_pattern}.
%%%%%%%%%%%%%%%%%%%%%%%%%%%%%%%%%%%%%%%%%%%%%%%%%%%%%%%%%%%%%%%%%%%%%%%%%%%%%%%%%%%%%5
\begin{figure*}[!htp]
    \centering
   \includegraphics[width=\textwidth]{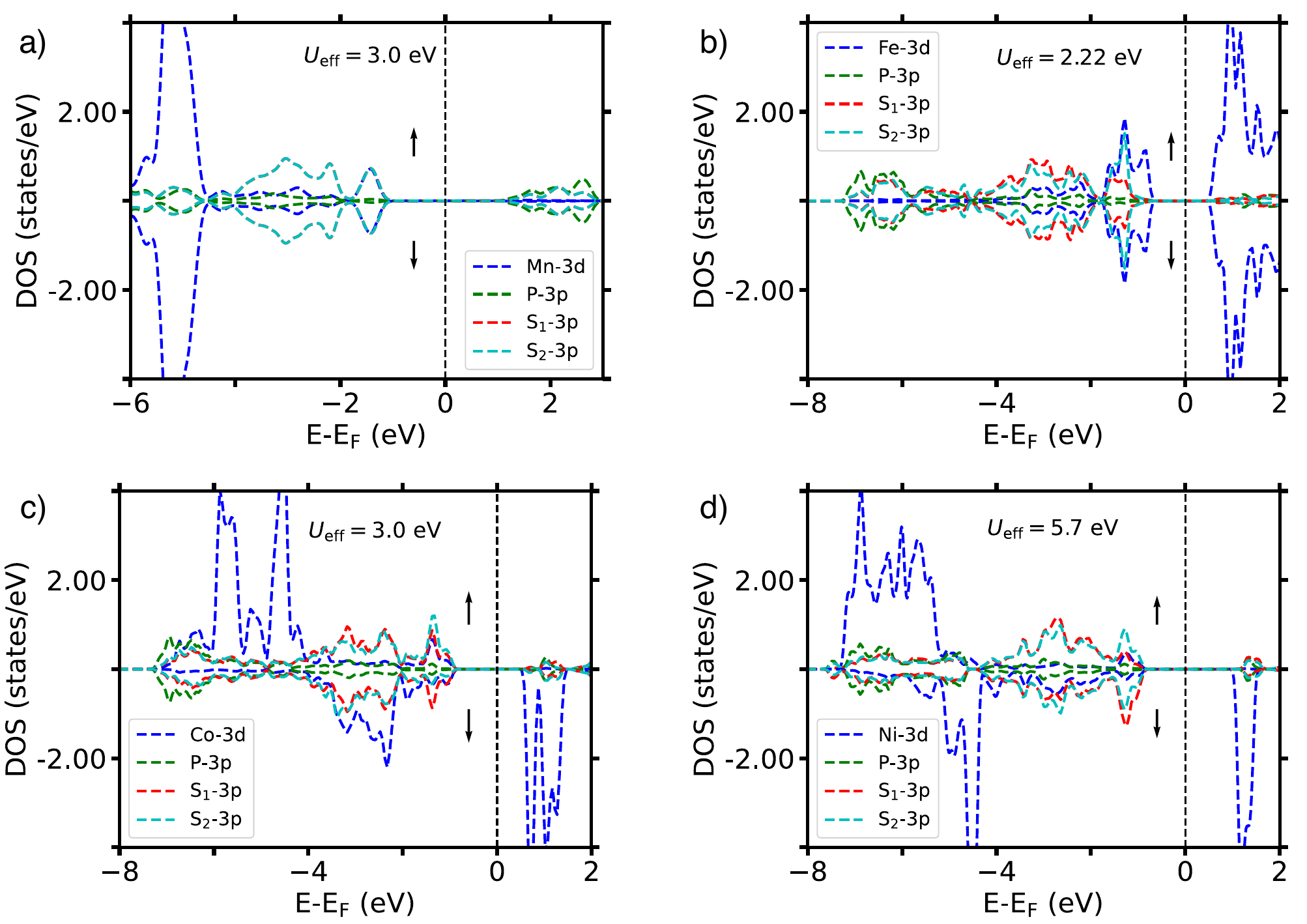}
\caption{(Color online) Electronic density of states (DOS) calculated with GGA+$U$ at optimized lattice
coordinates for various magnetic orderings: (a) N\'eel-AFM
(b) Long-bond zigzag (c) Short-bond zigzag.  For these calculations, we consider one atom of each species and plot the DOS for spin-up (up arrow) and spin-down (down arrow). The spin-up direction is defined by the majority spin of the magnetic ion selected for the plot. According to Wyckoff's positions, two types of S atoms have different distances from magnetic ions. For Mn, the distances Mn-S$_{1}$ and Mn-S$_{2}$ are 2.690 and 2.687 \AA, respectively. That's why the DOS for S-3p coincide with each other. For Fe, Fe-S$_{1}$ and Fe-S$_{2}$ are 2.616 and 2.609 \AA, respectively. For Co, Co-S$_{1}$ and Co-S$_{2}$ are 2.523 and 2.520 \AA, respectively. For Ni, Ni-S$_{1}$ and Ni-S$_{2}$ are 3.323 and 3.229 \AA, respectively.
}
\label{fig:DOS}
\end{figure*}
%%%%%%%%%%%%%%%%%%%%%%%%%%%%%%%%%%%%%%%%%%%%%%%%%%%%%%%%%%%%%%%%%%%%%%%%%%%%%%%%%%%%%%%%%%%%%%%%%%%%

Therefore, high-accuracy geometry optimization is crucial, requiring careful selection of the $U$ parameter and accurate representation of the magnetic order. 
For MnPS\(_{3}\), a primitive cell spanned by $a$ and $b'$ 
with two Mn ions is utilized, aligning with its N\'eel ground state. In contrast, for FePS\(_{3}\), CoPS\(_{3}\), and NiPS\(_{3}\), we employ both $2 \times 1$ and $1 \times 2$ supercells comprising four magnetic ions to adequately represent short-bond and long-bond zigzag states, respectively. The optimized lattice constants, presented in Tab.~\ref{tab1}, show excellent correlation with experimental values.
The $U$ parameter is meticulously chosen to enhance electron-electron correlation within the d-shells. Determining the optimal $U$ value proves challenging, as the value derived using the Density Functional perturbation theory \cite{Timrov2018} fails to reproduce the experimental bandgap. Consequently, we adjusted the $U$ parameter to align with the experimental bandgap. This adjustment also considers the variability in reported experimental bandgaps for bulk materials, adding to the complexity of accurately determining the $U$ value. The final $U$ parameters, which effectively describe the bandgap, structural properties, and magnetic moments of the ions, are detailed in Tab.~\ref{tab1}.
%##########################################################################################################

\begin{figure*}[!htp]
    \centering
   \includegraphics[width=\textwidth]{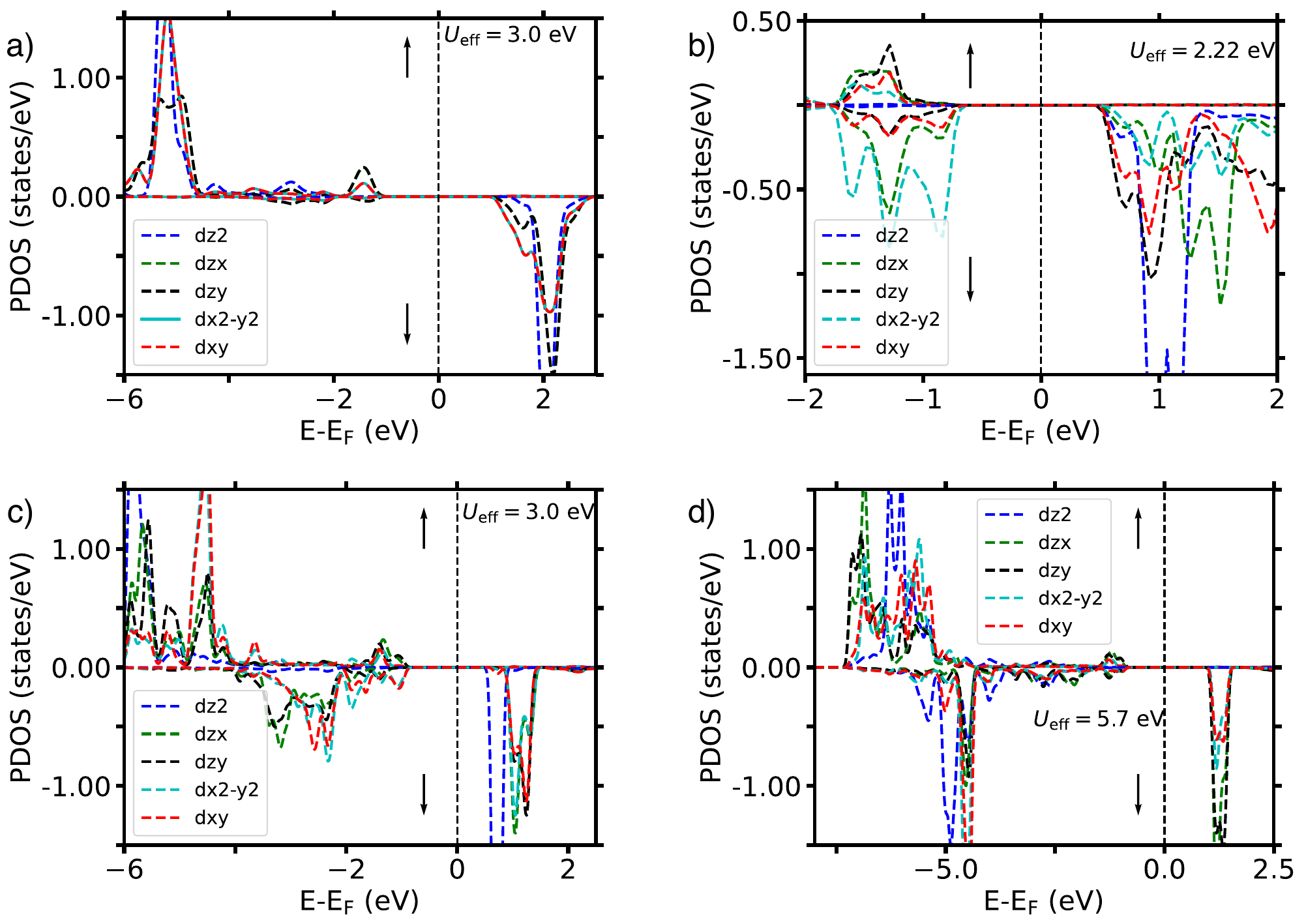}
\caption{(Color online) 
 Orbital-resolved electronic density of states for (a) MnPS$_3$, (b) FePS$_3$, (c) CoPS$_3$ and (d) NiPS$_3$ monolayers. We consider the ground-state spin pattern for each material. For these calculations, we consider one atom of each species and plot the DOS for spin-up (up arrow) and spin-down (down arrow). The spin-up direction is defined by the majority spin of the magnetic ion selected for the plot. }
\label{fig:PDOS}
\end{figure*}
%%%%%%%%%%%%%%%%%%%%%%%%%%%%%%%%%%%%%%%%%%%%%%%%%%%%%%%%%%%%%%%%%%%%%%%%%%%%%%%%%%%%%%%%%%%%%

Fig.~\ref{fig:DOS} in our study elucidates the contributions of p and d orbitals from P, S, and $M$ (Mn, Fe, Co, Ni) ions in $M$PS\(_{3}\) materials. For FePS\(_{3}\), in the energy range extending from the Fermi level to $-2$ eV, the d orbitals of Fe are predominantly influential, especially for the shallow states. In contrast, deeper energy bands demonstrate a pronounced hybridization between Fe's d-electrons and the p-electrons of sulfur. For MnPS\(_{3}\), CoPS\(_{3}\), and NiPS\(_{3}\), the p orbitals of sulfur exhibit a more pronounced effect below the Fermi energy. This distinct behavior in FePS\(_{3}\) can be attributed to the crystal field effects arising from its distorted honeycomb lattice, leading to orbital ordering where the in-plane d\(_{x^2-y^2}\) orbital plays a pivotal role. This analysis underscores the significant impact of crystal field and lattice structure on the electronic states of these materials.
%%%%%%%%%%%%%%%%%%%%%
\begin{table*}[!htp]
  \centering
  \caption{Calculated Heisenberg  couplings $J_i$ (meV) up to the forth neighbors, bi-quadratic exchange interaction $B$ (meV), Dzyaloshinskii-Moriya exchange interaction $D$ (meV) and single ion anisotropy $\Delta$ (meV) for different $U_\text{eff}$(eV) parameters. Negative and
    positive value denotes anti-ferromagnetic and FM
    exchange interaction, respectively.
    Note that $|S|=1$ has been used in the definition of the spin Hamiltonian. By using the obtained couplings, we perform MC simulations to find the N\'eel ($T_\text{N}$) temperature (K). The temperature value in parenthesis would result if the biquadratic couplings were neglected.}
      \begin{tabular}{cccccccccccc}
    \hline
    Material &$U_\text{eff}$ & $J_\text{1a}$ &$J_\text{1b}$ &$J_\text{2}$ & $J_\text{3}$& $J_\text{4}$ &$\Delta$&$D$&$B$&$T_\text{N}$\\
    \hline
    MnPS$_{3}$& 3.00 &  $-7.89$ & $-7.89$& $-0.21$  & $-3.47$ & $0.02$ & $-0.025$ & 0.00 & $-0.95$ & 76.2 (73.0) \\
    \hline
    FePS$_{3}$& 2.22 & $-3.26$& 4.01 & $-1.24$ & $-5.71$ & 1.50 &$-0.89$&$-0.34$&$-2.10$&70.0 (66.9) \\
     \hline
     CoPS$_{3}$& 3.00 & $3.47$ &$3.47$ & $0.64$  & $-10.85$ & $0.06$ &$-0.14$&$ 0.00$&$-5.53$& 86.5 (67.7)\\
        \hline
    NiPS$_{3}$& 5.70 & $2.46$& $2.46$ & $0.14$  & $-11.58$ & $0.06$ & $-0.22$ & $0.00$ & $-6.91$ & 94.0 (70.6) \\
        \hline
    \label{tab2}
  \end{tabular}
\end{table*}

\begin{figure}[!htp]
    \centering
\includegraphics[width=0.8\columnwidth]{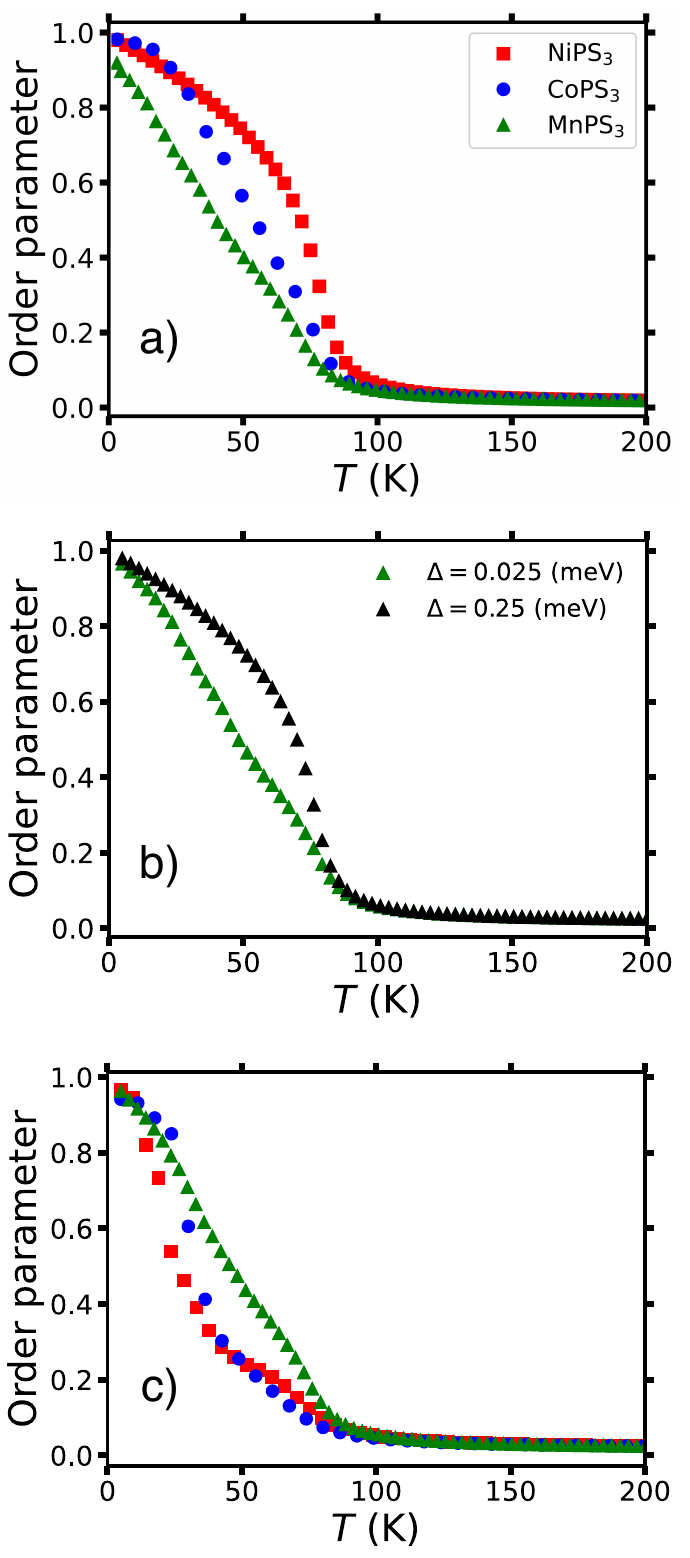}
\caption{(Color online) 
Order parameter versus temperature  (a,b) with, and (c) without bi-quadratic exchange interaction. Without the $B$-term, the critical temperature is by 73, 67.7, and 70.6 K for Mn, 
Co, and Ni, respectively.  (b) shows the effect of increasing $\Delta$ on the order parameter of MnPS$_{3}$. 
The order parameter \( m \) is defined as 
m $\equiv \frac{1}{N} \sum_{i} S_i \cdot \hat{d}$,
where \( S_i \) represents the direction of the magnetic moment at site \( i \), and \( \hat{d} \) represents the easy axis direction.
}
\label{fig:order_parameter}
\end{figure}

%########################

Fig.~\ref{fig:PDOS} depicts the orbital-resolved electronic density of states for MnPS\(_{3}\), FePS\(_{3}\), CoPS\(_{3}\), and NiPS\(_{3}\) monolayers, provides key insights into how the crystal field affects d-shell splitting. For MnPS\(_{3}\), CoPS\(_{3}\), and NiPS\(_{3}\), the d\(_{zy}\) and d\(_{zx}\) orbitals are particularly influential near the Fermi energy, a characteristic tied to t$_\text{2g}$ orbitals. In contrast, for FePS\(_{3}\) with its distorted lattice, the d\(_{x^2-y^2}\) (e$_\text{g}$) orbital becomes more significant. This difference underscores the impact of lattice geometry on electronic properties, particularly in how it influences the behavior of d-sublevels in these materials.  Since the magnetic properties in $M$PS$_{3}$ are governed by the super-exchange mechanism, the p-orbital of S ions as intermediate between magnetic ions plays a crucial role. According to Anderson's rule\cite{Anderson}, when $t$ term as kinetic energy in Hubbard Hamiltonian increases, the $J$ value will get stronger. In addition, the $t$ term is controlled by the level of hybridization. For MnPS$_{3}$, the hybridization is stronger as compared to other materials, that's why $J_{1}$ is larger than other materials (Tab.~\ref{tab2}).

Here, we study the effective spin Hamiltonian for $M$PS$_{3}$ monolayers.
In our investigation, we calculate the effective spin Hamiltonian, aiming to identify the ground state in large simulation cells and to analyze the finite-temperature properties of $M$ (= Mn, Fe, Co, Ni)PS\(_{3}\) monolayers. 
Fig.~\ref{fig:geometry} illustrates the various exchange interactions  according to distances between magnetic ions
and Tab.~\ref{tab2} reports their values (in meV, for $|S|=1$), as well as the optimum value of $U_\text{eff}$ parameter in eV used for each compound.
Notably, the geometry optimization performed at the start of our calculations shows that the distances between neighboring $M$ atoms are sufficiently distinct only in the case of FePS$_3$. This is why distinct first-neighbor interactions ($J_{1a}$ and $J_{1b}$) are determined only for Fe, whereas Mn, Co, and Ni can be described with one unique first nearest-neighbor exchange interaction.
Since $J_4$ is found to be considerably smaller than the other interactions, we can assume that the Heisenberg Hamiltonian is converged with respect to the interaction range considered. 
Interestingly, the parameter $J_3$ that connects parallel chains of the honeycomb lattice is always negative, i.e. AFM exchange is preferred. 
The absolute value of $J_3$ increases when going from Mn over Fe and Co to Ni, as already noticed earlier \cite{Autieri2022}. 
Thus, the tendency to AFM chain interactions is found to increase with d-band filling. 
The same trend is observed for the biquadratic coupling; it increases toward the end of the transition metal series. However, the values reported in Tab.~\ref{tab2} refer to effective spins normalized to $|S|=1$, while the size of the magnetic moment decreases along the transition metal series from Mn to Ni.
Therefore, the single-ion anisotropy is found to be the largest for Fe. 
As we pointed out earlier, we attribute this finding to the unusually large orbital magnetic moment in Fe \cite{Amirabbasi2023}. 

MnPS$_3$ is identified as an almost ideal AFM monolayer.
This is concluded from $J_1$ and $J_3$ being both negative, 
a characteristic of the N\'eel-type antiferromagnetism. 
In absolute terms, $J$\(_{1}\) is significantly larger for MnPS\(_{3}\) than the more long-ranged interactions. 
Thus, the behavior of MnPS\(_{3}\) is in line with the general expectations for magnetic insulators and makes this material distinct from the others. 
The small next-nearest neighbor interaction $J_2$ would prefer antiparallel coupling, but is frustrated in the N\'eel-type ground state.
In the remaining three materials, FePS$_3$, CoPS$_3$ and NiPS$_3$, the dominant role of $J_3$ results in a magnetic ground state formed by zig-zag chains that are coupled antiferromagnetically to each other. 
The spins along the chains are aligned ferromagnetically, which is favorable due to the positive values of $J_1$ for Co and Ni. 
While in CoPS$_3$ and NiPS$_3$ the spin chains may run in any of the three directions compatible with the honeycomb lattice, 
two specific directions of the spin chain relative to the crystal lattice are selected by the distinct values of $J_{1a}$ and $J_{1b}$: 
the 'long bond' between two neighbor Fe atoms is part of the FM chain, as $J_{1b}$ favors FM interaction. 
The spin orientation alternates from one chain to the next (as in 
CoPS$_3$ and NiPS$_3$) to satisfy the antiparallel inter-chain coupling dictated by $J_3$ in all transition metal phosphosulfides considered here.

It is widely recognized that the mere consideration of bilinear Heisenberg exchange interactions falls short in accurately describing the magnetic behavior of complex materials \cite{Kartsev2020}. In scenarios devoid of spin-orbit coupling, the most consequential higher-order term emerges as the bi-quadratic term, delineated as the fourth-order perturbation within the framework of the Hubbard model \cite{Hoffmann_2020}. Notably, a positive \(B\) term predominantly facilitates the emergence of noncollinear spin configurations; conversely, a negative \(B\) value is instrumental in engendering collinear ground states. For the family of $M$PS$_{3}$ materials, empirical evidence substantiates the collinear nature of the magnetic ground states, corroborated by the observation of negative \(B\) terms. These findings are systematically documented in Tab.~\ref{tab2}.
Specific attention is bestowed upon FePS$_3$, owing to the proximal spatial arrangement of its first nearest neighbors, necessitating the calculation of both ($B_\text{1a}=-$2.10 meV) and ($B_\text{1b}=-$1.22 meV). 
To elucidate the influence of the \(B\) term on the critical temperature and order parameter of $M$PS$_3$ materials, MC simulations were meticulously conducted with \(B\) set to zero. The resultant critical temperatures, adjusted in light of the removal of \(B\), are presented in parentheses in Tab.~\ref{tab2}. The omission of the \(B\) term precipitates a diminution in the critical temperature, attributable to the attenuation of exchange coupling's capacity to counterbalance thermal fluctuations.

The presentation of the results concerning the order parameters of N\'eel and short-bond zigzag orders, as derived from the spin Hamiltonian (Eq.~\ref{H}), is illustrated in Fig.~\ref{fig:order_parameter}. It is noted that the definition of the order parameter for long-bond zigzag order in FePS$_{3}$ is not feasible through classical MC simulations without incorporating spin-phonon coupling~\cite{Amirabbasi2023}. For the elements Mn, Co, and Ni, the observed transitions exhibit characteristics akin to those of the Kosterlitz-Thouless transition, attributed to the relatively weak single-ion anisotropy. Conversely, FePS$_{3}$ demonstrates a strong out-of-plane easy axis anisotropy, aligning its behavior more closely with that predicted by the Ising model.
The Monte Carlo simulations underscore the insufficiency of the $B$-term alone to significantly alter the order parameter (or the magnetic ground state spin pattern). However, they do indicate 
a significant increase of the critical temperature if the $B$-term is included.

Regarding the calculated values for single-ion anisotropy, these are detailed in Tab.~\ref{tab2}. The necessity of employing GGA+$U$+SOC for these calculations is highlighted, emphasizing the pivotal role of the orbital moment. As summarized in Tab.~\ref{tab3}, which categorizes each system's spin and orbital moments as per GGA+$U$+SOC calculations, Fe exhibits the largest orbital moment, thus signifying considerable $\Delta$. This is plausible given Fe's electronic configuration ending at 3d$^{6}$. Conversely, Mn, with a closed-shell configuration (3d$^{5}$), exhibits an almost negligible orbital moment.
The smallness of the energy scale associated with $\Delta$ is also evidenced by the possibility of a spin-flop transition \cite{NanoLett.Morpurgo2020,PRL.Matthiesen2023} in this material.
To elucidate the orientation of the easy axis for each material, we analyze the lattice vectors as depicted in Fig.~\ref{fig:geometry}, calculating the total energies utilizing the GGA+$U$+SOC method. Our calculations confirm that for Fe-based systems, the easy axis aligns with the $c$-direction, perpendicular to the $a-b$ plane. In contrast, for Co and Ni, the easy axes are oriented along the $b$ and $a$ directions, respectively. In the case of Mn, 
there is a slight preference of orienting the spins in $b$ direction, i.e. along Mn--Mn bonds, but the energy differences are so small that one may likewise speak of 
the $a-b$ plane as an easy plane.
Another consequence of SOC is the Dzyaloshinskii-Moriya interaction. As indicated in Tab.~\ref{tab2}, the
$D$ term is nonzero exclusively for Fe. This is attributed to the non-ideal honeycomb lattice structure of Fe-based systems, which lack inversion symmetry. Conversely, the other materials, characterized by an ideal honeycomb lattice, exhibit inversion symmetry, resulting in a zero 
$D$ term.

Table~\ref{tab2} also includes the calculated critical temperatures (in K) obtained from Monte Carlo simulations. 
At low temperatures, these simulations converge to the AFM ground state of the respective material, i.e. N\'eel in MnPS$_3$ and chain-like in the other three compounds. From FePS$_3$ over CoPS$_3$ to NiPS$_3$, we find a trend towards increasing ordering temperature $T_N$. 
Fig.~\ref{fig:order_parameter} illustrates the order parameter of the magnetic ground state for $M$PS$_{3}$. Following the Mermin-Wagner theorem \cite{Mermin1966}, the absence of anisotropic exchange interactions precludes the possibility of a thermodynamically stable phase transition and enduring magnetic order in two-dimensional systems. However, in this instance, the order parameter asymptotically approaches unity at low temperatures, yet exhibits an abrupt decline to zero, indicating a lack of well-defined stability. To address this, the strength of single-ion anisotropy of MnPS$_{3}$ was incrementally increased from 0.025 to 0.25~meV (Fig.~\ref{fig:order_parameter}b), analogous to the application of a magnetic field along the easy-axis direction. Consequently, it can be inferred that MnPS$_3$, in its two-dimensional form, maintains a stable N\'eel ground state at low temperatures. To enhance this stability at elevated temperatures, the application of a magnetic field or an increase in the influence of single-ion anisotropy is necessitated.

\subsection{Magnon spectra}

Starting from the spin Hamiltonian, Eq.~(\ref{H}), we calculate magnon spectra using the Holstein-Primakoff transformation and linearizing around the magnetic ground state of each compound. 
In this procedure, the biquadratic term in the Hamiltonian is considered approximately via a renormalization of the nearest-neighbor Heisenberg couplings and the on-site anisotropy constant. For discussing the magnon spectra of FePS$_3$, CoPS$_23$ and NiPS$_3$, we use the rectangular magnetic unit cell spanned by $a$ and $b$ that also forms the basal plane of the monoclinic cell used in the literature. 
In the Brillouin zone, the reciprocal-lattice directions $a^* \parallel \Gamma X$ and $b^* \parallel \Gamma Y$ are perpendicular to $b$ and $a$, respectively. 
More details of the calculations can be found in the appendix.

Results for MnPS$_3$ are shown in Fig.~\ref{fig:MnPS3magnons}. For the N\'eel ground state, the crystallographic 
and the magnetic unit cell are identical (both hexagonal honeycomb lattice), and magnon dispersions are shown along the $\Gamma K$ and $\Gamma M$ path. 
Since both the single-ion anisotropy and the biquadratic term are small for this material, the zero-energy gap in the magnon spectrum is tiny (barely visible in the plot).
Our result for the magnon spectrum of MnPS$_3$ can be compared with the spectra calculated by Olsen \cite{Olsen} and by Bezazzadeh {\em et al.}\cite{Bezazzadeh2021}. 
In their calculations, the magnon spectra reach their maxima at about 8~meV if $U=5$ eV is used\cite{Olsen,Bezazzadeh2021}, and about 13~meV  \cite{Olsen}, similar to ours, using $U=3$ eV.  This confirms the commonly observed trend that large $U$ leads to weaker magnetic interaction, and hence softer magnon spectra.

\begin{figure}[htb]
\centering
\includegraphics[width=\columnwidth]{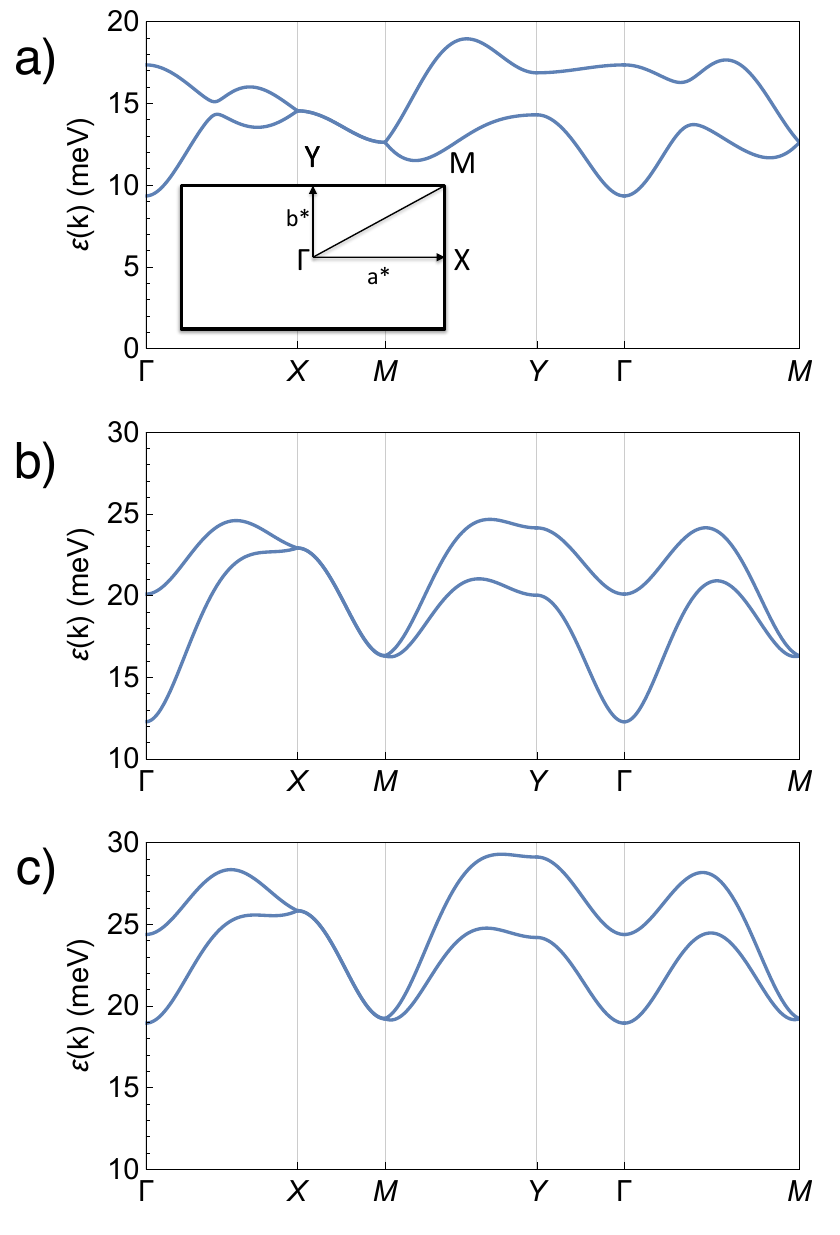} 
\caption{\label{fig:magnons}  Magnon spectra for (a)  FePS$_3$ in the long-zigzag chain AFM state, (b) CoPS$_3$ in the short-zigzag chain AFM state, (c) NiPS$_3$ in the short-zigzag chain AFM state. In the inset, the path in the Brillouin zone is shown. 
For all materials, zig-zag chains of parallel spins run along $\Gamma X$.}
\end{figure}

For FePS$_3$, CoPS$_3$ and NiPS$_3$ that possess zig-zag chains as their magnetic ground state the magnetic unit cell is twice as large as the crystallographic unit cell, i.e., it contains four transition metal atoms. 
Consequently, the magnon spectra of these materials, shown in Fig. \ref{fig:magnons}, develop two branches. 
The lower branch does not reach zero at the small wavevector, as one would expect for 'acoustic' FM magnons; the sizeable gap in the magnon spectra at $\Gamma$ even increases in size when going from Fe to Co to Ni. 
From our calculated exchange interactions, we conclude that the increasing biquadratic coupling term is mostly responsible for opening this gap.
We note that such a gap is known to give rise to a logarithmic correction to the magnetic ordering temperature, see e.g. Ref. \onlinecite{Kartsev2020}.
The upper and the lower magnon branches show large splitting along $\Gamma Y$ which increases from Fe to Ni. 
This dispersion reflects the AFM coupling between the chains of parallel spin; its size is mostly governed by the exchange constant $J_3$ that increases along the transition metal sequence, as evidenced by the data in Tab. \ref{tab2}. 
The direction $\Gamma X$ in the spectra reflects the dispersion {\it along} the chains of parallel spin. 
The spectra for FePS$_3$, the material with the 'long-bond' zigzag ground state, shows marked difference to CoPS$_3$ and NiPS$_3$ that have isotropic nearest-neighbor exchange $J_{1a} = J_{1b}$. 
In FePS$_3$, one observes an avoided crossing of the two magnon branches along $\Gamma X$. This occurs because the interactions $J_{1a}$ and $J_{1b}$ that couple the parallel spins in the zig-zag chain have opposite sign. 
As a consequence, the energy of the upper branch at $\Gamma$ goes below the lower branch when reaching $X$, and vice versa for the lower branch starting at $\Gamma$ that rises in energy.

\begin{figure}[tbh]
\centering
\includegraphics[width=0.8\columnwidth]{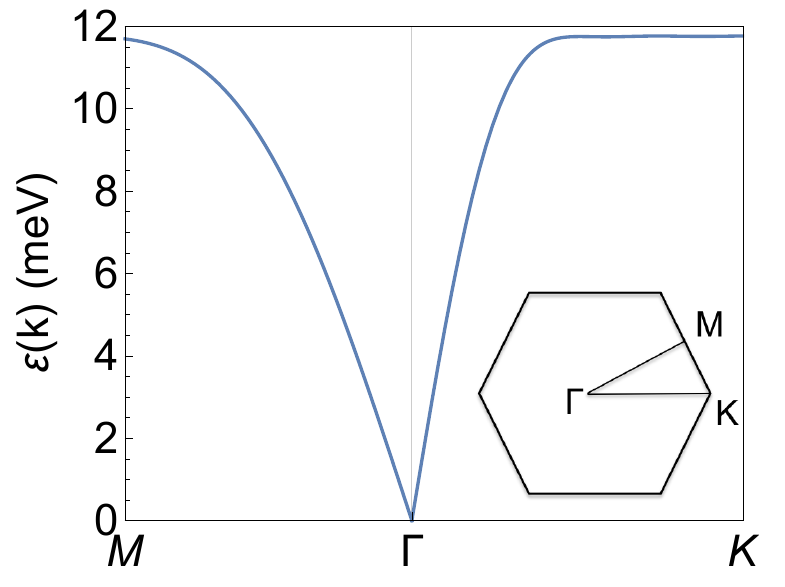}
\caption{\label{fig:MnPS3magnons}  Magnon spectrum for MnPS$_3$ in the N\'eel state. In the inset, the path in the Brillouin zone is shown.}
\end{figure}

Recently Yan {\em et al.}\cite{Yan2024} calculated magnon spectra {\em in bulk samples} for all four compounds studied here on the 
basis of a bi-linear Hamiltonian with exchange interactions up to the third neighbor shell. 
This means the forth neighbor interactions were not taken into consideration; moreover, they found a relatively 
large single-ion anisotropy but ignored biquadratic interactions. The range of dispersion of the magnons predicted 
by them is in reasonable agreement with our magnon spectra, although details are different. 
This could be due to the difference in the magnetic ground state used as the starting point (Yan {\em et al.} did not distinguish between 'long-bond' and 'short-bond' zig-zag chains), 
or due to differences between the bulk and the monolayer.

Finally, we compare our calculated magnon spectra of monolayers to experimentally observed spectra from inelastic neutron scattering at bulk materials. 
For MnPS$_3$, we find overall good agreement with the experimental spin wave dispersion published in Ref.~\onlinecite{Wildes1998}.
In this inelastic neutron scattering study, the spin waves reach their maximum energy at 11.5~meV, very close to our result of 12 meV.
The observed magnon gap at the $\Gamma$-point is small, less than 0.5~meV, thus confirming the very small value of the magnetic anisotropy found in our calculations. 
Moreover, a recent neutron spectroscopy study \cite{Wildes_2021} showed that the DMI in this material is negligibly small, which matches with our calculations.

For FePS$_3$, the inelastic neutron scattering data \cite{Lancon2016} shows a magnon branch starting at 17~meV dispersing downward 
to about 15~meV and then bending up again. Because of the incipient downward dispersion, we believe that this is the upper branch in our calculated spectrum,
whereas the lower branch was not observed. 
In a later analysis of the experimental data by the same group \cite{Wildes2020}, an improved fit of the data has been obtained by invoking a biquadratic coupling, albeit with a smaller exchange constant $K$ than calculated by us. 
The experimental fit resulted in a very large single-ion anisotropy of about 2.5 meV to explain the magnon gap, whereas our theoretical description works with a much smaller $\Delta =- 0.89$meV and explains the magnon gap by a relatively large value of $K$ which effectively renormalizes the anisotropy. 
Moreover, we speculate that the branches at higher energy (up to 40~meV) detected in the neutron scattering 
experiment are mixed phonon-magnon branches with a small magnon admixture, since according to our calculations FePS$_3$ 
does not support such hard pure magnon modes. 
A hybridization of magnon and phonon excitations in FePS$_3$ has been proposed recently \cite{Vaclavkova2021,Ghosh2021,To2023} on theoretical grounds.

For CoPS$_3$, inelastic neutron scattering \cite{Wildes_2017,Wildes_2023} detected magnonic losses at 15~meV and 33~meV. The lower value is in the range where our calculation predicts a magnon branch. 
The experimentally observed upper branch might again result from a hybridization with phonons.

For NiPS$_3$, inelastic neutron scattering \cite{Lancon2018} detected losses both below 10~meV and in the 40 to 50 meV range. Magnon models fitted to experimental inelastic neutron scattering data \cite{Lancon2018,Wildes_2020NiPS3} placed the magnon bands in the range of 8~meV, dispersing up to 50~meV. This is rather different from the magnon spectra presented here, which start at higher energy  (18~meV) but then show less dispersion, reaching up to 28~meV. The reason for the disagreement is presently not understood.     
However, both magnon branches of NiPS$_3$, similar to those of CoPS$_3$, show strong disperions along the $\Gamma Y$ direction. This indicates the strong AFM interaction between chains encoded in the large (and negative) parameter $J_3$, which is in qualitative agreement with the experimental findings.

\section{Conclusion}
\label{sec:conclusion}
In conclusion, this study offers a detailed examination of the magnetic properties of $M$PS$_{3}$ 2D materials, utilizing the DFT+$U$+SOC approach and Monte Carlo simulations. It highlights significant findings in the understanding of Heisenberg couplings, biquadratic and Dzyaloshinskii-Moriya interactions, and single ion anisotropy across various $M$PS$_{3}$ compounds. Since these materials are 3D magnetic semiconductors, applying the Hubbard parameter is essential to enhance the electron-electron correlations. The optimum \( U \) parameters have been chosen as 3.0 eV for Mn, 2.22 eV for Fe, 3.0 eV for Co, and 5.57 eV for Ni, respectively. Although magnetic ions form a hexagonal lattice in all materials, the way electrons fill the d shells of these ions results in different spin patterns in the ground state. 
The geometry optimization reveals that only in the case of FePS$_{3}$, there is a distortion from the ideal hexagonal lattice configuration. Specifically, the Fe-Fe distances between nearest neighbors vary by 0.14\AA. 
Due to this distortion, we calculate two different biquadratic exchange terms along the longer and shorter bond distances. 

To clarify the importance of the biquadratic exchange for these materials, we conduct Monte Carlo simulations both with and without the biquadratic term. Our results indicate that the absence of the biquadratic term leads to a reduction in the N\'eel temperature and alters the order parameter, which signifies the spin-spin correlation. Specifically, the order parameter undergoes a transition from 1 (indicating long-range order) to 0 (indicating a paramagnetic phase) at lower temperatures.
In MnPS$_{3}$, this effect is less pronounced due to a smaller biquadratic term. MnPS$_{3}$, with its N\'eel ground state, behaves like a typical magnetic semiconductor where the exchange interaction parameters ($J$) decrease with increasing distances. 
Additionally, the N\'eel temperature obtained from MC simulations is 73.6 K, which closely aligns with the experimental value of 78 K observed in the bulk system. However, for other materials that feature AFM zigzag chains, the third-nearest-neighbor exchange interaction ($J_{3}$), which connects these chains, is crucial for stabilizing the zigzag ground state, especially in the case of NiPS$_{3}$.
Spin-orbit coupling effects, particularly single-ion anisotropy and Dzyaloshinskii-Moriya interactions, along with orbital moment (0.77 $\mu_\mathrm{B}$), are most pronounced in Fe due to the electronic configuration of Fe\(^{2+}\) which ends up as \(3d^6\). For other materials, the Dzyaloshinskii-Moriya interaction values are nearly zero, and the orbital moments are 0.02, 0.11, and 0.22 $\mu_\mathrm{B}$ for Mn, Ni, and Co, respectively. Consequently, the strength of the single-ion anisotropy in these materials follows a similar trend to that of the orbital moments.
While the effect of biquadratic interactions on the gaps in the magnon spectra is sizeable, the increase in critical temperatures ranges from 5\% to $\sim 30$\% from early to late transition metal ions.
\section{ACKNOWLEDGMENT}
We gratefully acknowledge the computing time granted by the Center for Computational Sciences and Simulation (CCSS) of the University of Duisburg-Essen and provided on the supercomputer magnitUDE (DFG Grant No. INST 20876/209-1 FUGG and INST 20876/243-1 FUGG) at the Zentrum f\"ur Informations- und Mediendienste (ZIM). M. A. was supported by a fellowship from Universit\"at Duisburg-Essen.

\appendix 
\section{Derivation of spin Hamiltonian}
\paragraph{Derivation of Heisenberg term}
The exchange parameters of the Heisenberg model as detailed in Eq.~(\ref{eq:S-1}) are determined by fitting to total energies obtained for the ions fixed at their previously determined, materials-specific ground state positions, but varying collinear spin configurations, similar to our previous work~\cite{Amirabbasi2023}. 
We analyze a supercell configuration of dimensions 2$\times$2, comprising 40 ions that include 8 magnetic atoms, alongside 8 phosphorus  and 24 sulfur ions. Analytical calculations permit the determination of distances between magnetic atoms extending to the fourth-nearest neighbors. Each category of nearest neighbor interaction is associated with a corresponding exchange coupling constant, denoted as $J$. It is pertinent to mention that in our analytical derivations, the magnitude of the spin, $|S|$, is set to 1, which simplifies the calculation of the exchange parameter $J$.

Prior to conducting Density Functional Theory augmented with Hubbard $U$ (DFT+U) calculations, up to 15 collinear spin configurations are postulated, and their total energies are computed analytically. Here, ferromagnetic (FM) interactions among nearest neighbors contribute a value of $+1/2$, whereas antiferromagnetic (AFM) interactions contribute a value of $-1/2$. Subsequently, the total energies for these configurations are quantitatively assessed using the DFT+U method. By employing a least-squares fitting technique,  the low-lying energies are mapped onto the previously derived analytical expressions, thereby facilitating the extraction of the $J$ parameters.
\begin{eqnarray}
H_{\rm {Heis}}=-\sum_{i\neq j} J_{ij}(\vec{S_{i}}\cdot\vec{S_{j}})
\label{eq:S-1}
\end{eqnarray}

\paragraph{Derivation of biquadratic term}
To derive the biquadratic exchange term, $B$, it is necessary to analyze noncollinear configurations where the contributions from the Heisenberg exchange term in the total energy differences are degenerate. To achieve this, we consider a 2$\times$1 supercell configuration, comprising 20 atoms with 4 magnetic ions. The spins of the first and last magnetic ions are held fixed, while the orientations of the spins of the intervening magnetic ions are rotated such that the sum of their spins equals zero, i.e., $S_i + S_j = 0$. This configuration ensures that the observed variations in total energy are solely attributable to the biquadratic term, $B$.
For the specific case of FePS$_{3}$, to calculate the biquadratic term $B_{1b}$, we adopt a similar strategy but utilize a 1$\times$2 supercell. Here, the spatial arrangement of the atoms within the supercell is designed such that the distances between the intermediate atoms correspond to $d_{1b}$. This approach enables the isolation of the biquadratic exchange contributions from other magnetic interactions within the system.
\begin{eqnarray}
H_{\rm {B}}=B\sum_{\rm n.n} (\vec{S_{i}}\cdot\vec{S_{j}})^{2}
\label{eq:S-2}
\end{eqnarray}

\paragraph{Derivation of anisotropic terms}
For the evaluation of anisotropic terms, all calculations are conducted using the primitive cell, which comprises 10 atoms with 2 magnetic ions. 
To deduce the strength of the Dzyaloshinskii-Moriya interaction (DMI), it is essential to consider the influence of spin-orbit coupling (SOC). The computational framework utilized for these calculations is GGA+$U$+SOC. Within this framework, we need to select magnetic configurations in such a way that the differences in total energies primarily reflect the effects of SOC, while the contributions from single-ion anisotropy after SOC and other interactions before SOC are rendered degenerate.
We examine two distinct magnetic configurations. In the first configuration, the orientations of the spins are aligned along the '$a$' axis and the negative '$b$' axis. In the second configuration, the spins are aligned along the '$a$' axis and the positive '$b$' axis.  The total energy differences of these two configurations can be attributed to DMI.

For the determination of single-ion anisotropy (SIA), we analyze a pair of magnetic configurations wherein all spins are aligned with either the easy-axis or the hard-axis. These orientations are strategically chosen to ensure the cancellation of the DMI within the framework of GGA+$U$+SOC calculations. Consequently, the differences in total energies observed between these configurations can be exclusively attributed to the effects of SIA. It should be noted that the total energies of the two configurations are the same before considering SOC, indicating the degeneracy of isotropic terms.
\begin{eqnarray}
H_{\rm {Anisotropic}}=D \sum_{\rm n.n} \hat{D}_{ij}\cdot(\vec{S_{i}}\times \vec{S_{j}})+\Delta\sum_{i} (\vec{S_{i}}\cdot\vec{d_{i}})^{2}
\label{eq:S-3}
\end{eqnarray}

\section{Calculation of magnon spectra}
\paragraph{Bosonisation of spin operators}
We define spin operators $\mathbf{\hat S} = (\hat S^x, \hat S^y, \hat S^z) =  \bigl( \frac{1}{2}(\hat S^+ + \hat S^-), \frac{1}{2}(\hat S^+ - \hat S^-), \hat S^z \bigr)$ at lattice site $j$ via bosonic creation and annihilation operators
\begin{eqnarray*}
\hat S_j^+ &=& \sqrt{2S - \frac{ \hat b^\dagger_j \hat b_j }{2S} } \, \hat b_j  \, , \\
\hat S_j^- &=& \hat b_j^\dagger \sqrt{2S - \frac{ \hat b^\dagger_j \hat b_j}{2S} } \, , \\ 
\hat S^z_j &=& S - \hat b^\dagger_j \hat b_j 
\end{eqnarray*}
For sufficiently large $|S|$, it is sufficiently accurate to use a Taylor expansion of the square root to lowest order, yielding
\begin{eqnarray*}
\hat S_j^+ &\approx& \sqrt{2S} \, \hat b_j  \, , \\
\hat S_j^- &\approx& \hat b_j^\dagger \sqrt{2S} \, , \\ 
\hat S^z_j &=& S - \hat b^\dagger_j \hat b_j \, .
\end{eqnarray*}

%\begin{minipage}[l]{0.65\textwidth}
For a detailed calculation of the Hamiltonian in terms of the $\hat b$ and $\hat b^\dagger$ operators, we refer to Kartsev {\it et al.}, Ref.~\onlinecite{Kartsev2020}, supplementary section~15. 
A Fourier transformation of the operators $\hat b$ and $\hat b^\dagger$ is performed for each inequivalent lattice site, and consequently the operators acquire an additional index $k$. 

In a honeycomb lattice, there are two inequivalent lattice sites, and if the Neel ground state is considered, these two sites are sufficient.  
For those systems that display zig-zag chains as their ground state, we use the magnetic unit cell containing four transition metal atoms, as shown in the figure. 
There are two chains of ferromagnetically aligned spins, 1--4--1--4\ldots and {2--3--2--3\ldots}, with spins between chains antiparallel to one another. 
Note that for FePS$_3$ the ferromagnetic chains consist of alternating long and short bonds, i.e. the bonds 2--3 and 4--1 are long ones, described by the interaction $J_{1b}$.  
The Hilbert space of the spins of the four atoms is spanned by the vector 
%\end{minipage}
%1\begin{minipage}[r]{0.3\textwidth}

\begin{figure}[htbp]
\begin{center}
\hspace{1cm} \includegraphics[width=3.5cm]{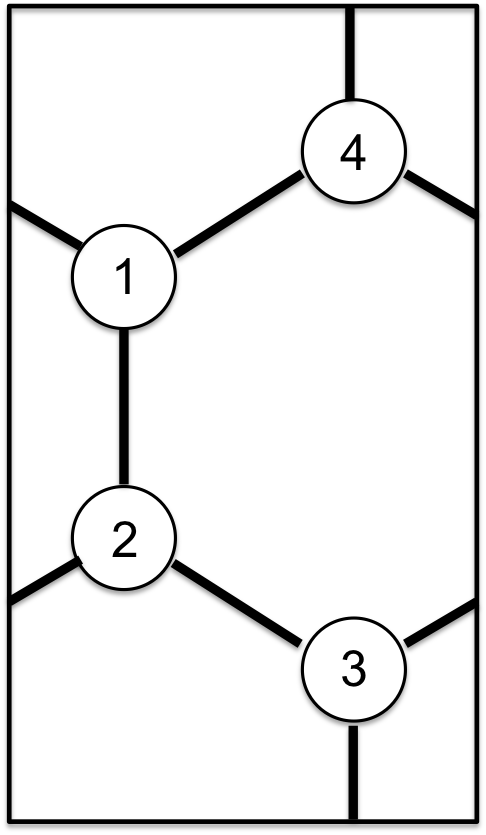} 
\caption{Supercell used to calculate magnons of materials with ferromagnetic zigzag chains as ground state. 
There are two chains, 1--4--1--4 \ldots and 2--3--2--3 \ldots, that are antiparallel to one another.}
\label{fig:S1}
\end{center}
\end{figure}
%\end{minipage}

$$
\Psi_\mathbf{k} = \left( b_{1,\mathbf{k}}, b^\dagger_{3,-\mathbf{k}}, b_{4,\mathbf{k}}, b^\dagger_{2,-\mathbf{k}}, b^\dagger_{1, -\mathbf{k}}, b_{3,-\mathbf{k}}, b^\dagger_{4,-\mathbf{k}}, b_{2,\mathbf{k}} \right) 
$$
The first indices of the operators refer to the atoms numbered in Fig. S1. Note that, in case of FePS$_3$, the long bond (with interaction $J_{1b}$) is between atoms 2 and 3.

\paragraph{Renormalization of interactions due to biquadratic term}

The biquadratic term, coupling two neighboring atomic sites $i$ and $j$, is approximated by
\begin{equation}
(\mathbf{S}_i \cdot \mathbf{S}_j )^2 \approx 2 S^3 (b^\dagger_i b_j + b_i b^\dagger_j - 3 b^\dagger_i b_i - 3 b^\dagger_j b_j) + S^4 \, ,\
\label{eq:approxB}
\end{equation}
retaining the highest powers in $|S|$.

The Hamiltonian $H_{\rm spin}$ in eq. (1) of the main text has on-site and nearest-neighbor terms, in addition to the more long-ranged terms in the Heisenberg Hamiltonian. The Dzyaloshinskii-Moriya interaction between nearest neighbors is very small; in fact it vanishes due to symmetry if all nearest neighbor bonds have the same length. Therefore we can disregard the canting of spins. 

Both the biquadratic term and the Heisenberg terms $J_{1a}$ and $J_{1b}$ run over nearest neighbors; using the approximation (\ref{eq:approxB}) it is therefore possible to merge them by introducing
\begin{eqnarray*}
\tilde J_{1a} &=& J_{1a} + B S^2 \\
\tilde J_{1b} &=& J_{1b} + B S^2 \\
\tilde \Delta &=& \Delta + 3B S^2
\end{eqnarray*}
Note that $\hat S_z$ invokes an on-site operator product $b_i^\dagger b_i$; this is why the on-site anisotropy $\Delta$ needs to be renormalized, too. The factor 3 appearing in the above equation stems from the 3 nearest neighbors. 

With these transformations, the Hamiltonian is brought to the bilinear form
\begin{equation}
H_{\rm spin} = - \frac{1}{2} \sum_\mathbf{k} \Psi^\dagger_{\mathbf{k}} H(\mathbf{k}) \Psi_{\mathbf{k}} + H_0
\label{eq:bilinHam}
\end{equation}

\paragraph{Holstein-Primakoff Hamiltonian}

In the following, we refer to the renormalized case with $|S|=1$, as in the main text of the manuscript. 
For zig-zag chains, a supercell containing four magnetic ions, Fig. S1,  is used. 
For notational brevity, we rescaled the $\mathbf{k}$-vector to match the square-shaped Brillouin zone, 
$k_x a \mapsto k_x, \quad k_y a \sqrt{3} \mapsto k_y$ with $a$ being the lattice constant.
After the bosonisation of the four spin operators, the Hamiltonian can be expressed by a $(8 \times 8)$ matrix (see Ref.~\onlinecite{Lee2018})

%\begin{equation}
$$
H(\mathbf{k}) = \left( \begin{array}{cc}  h(\mathbf{k}) & 0 \\ 0 & h^T(-\mathbf{k}) \end{array} \right) 
$$
%\end{equation}
The two blocks on the diagonal are given by $(4 \times 4)$ matrices $h(\mathbf{k})$ with 
%\begin{equation}
$$
h(\mathbf{k}) = \left( \begin{array}{cc}  d(\mathbf{k}) & \gamma(\mathbf{k}) \\ \gamma^\dagger(\mathbf{k})  & d(-\mathbf{k}) \end{array} \right) 
$$
%\end{equation}

For the long zig-zag phase, e.g. in FePS$_3$, the entries are expressed using Pauli's $(2 \times 2) \mathbf{\sigma}$-matrices
\begin{widetext}
\begin{eqnarray*}
d(\mathbf{k}) &=& \mathbf{\sigma_0} ( \tilde \Delta - \tilde J_{1b} + 2 J_2 + 3 J_3 -2 J_4 - 2 D \sin k_x +2 J_2 \cos k_x ) \\
 &+& \mathbf{\sigma_1} \left[ 4 \cos(k_y/2) ( D \sin(k_x/2) + J_2 \cos(k_x/2) )  \right]  \, \\
 \gamma(\mathbf{k}) &=& \mathbf{\sigma_0} \left[ (\tilde J_{1a}+ \tilde J_{1b}) e^{i k_y/6} \cos(k_x/2) + 2 J_4 \bigl( e^{-5 i k_y/6} \cos(k_x /2) + e^{ik_y/6} \cos(3 k_x/2) \bigr) \right] \\   
 &+ & \mathbf{\sigma_1}  \left[  2 J_3 \cos k_x + J_3 e^{i k_y} +2 J_4 e^{i k_y} \cos k_x  + \tilde J_{1a} \right] e^{-i k_y/3}
\end{eqnarray*}

For the short zig-zag phase, e.g. in CoPS$_3$ and NiPS$_3$, we have 
\begin{eqnarray*}
d(\mathbf{k}) &=& \mathbf{\sigma_0} ( \tilde \Delta - 2 \tilde J_{1a} + \tilde J_{1b} + 2 J_2 + 3 J_3 -2 J_4 - 2 D \sin k_x +2 J_2 \cos k_x ) \\
 &+& \mathbf{\sigma_1} \left[ 4 \cos(k_y/2) ( D \sin(k_x/2) + J_2 \cos(k_x/2) )  \right]  \, \\
 \gamma(\mathbf{k}) &=& \mathbf{\sigma_0} \left[ 2 \tilde J_{1a}  e^{i k_y/6} \cos(k_x/2) + 2 J_4 \bigl( e^{-5 i k_y/6} \cos(k_x /2) + e^{ik_y/6} \cos(3 k_x/2) \bigr) \right] \\   
 &+ & \mathbf{\sigma_1}  \left[  2 J_3 \cos k_x + J_3 e^{i k_y} +2 J_4 e^{i k_y} \cos k_x  - \tilde J_{1b} \right] e^{-i k_y/3}
\end{eqnarray*}
\end{widetext}

To calculate the magnon spectrum, we use the method described in Ref.~\onlinecite{Bezazzadeh2021}, 
implemented in {\sc Mathematica12}: %\cite{Mathematica}: 
First, we calculate the Cholevsky decomposition of $h(\mathbf{k})$, i.e. we calculate the square matrix $R(\mathbf{k})$ with the property $h(\mathbf{k}) = R^\dagger(\mathbf{k}) R(\mathbf{k})$. 
A matrix with both positive and negative eigenvalues is constructed by $X(\mathbf{k}) = R(\mathbf{k}) \hat \phi R^\dagger(\mathbf{k})$, where $\hat \phi$ is the so-called para-unitary matrix, i.e., a diagonal matrix with alternating entries of $1$ and $-1$, $(\hat \phi)_{ij} = (-1)^j \delta_{ij}$. 
The positive eigenvalues $\varepsilon(\mathbf{k})$ of $X(\mathbf{k})$ can be interpreted as the frequencies of the magnon branches; 
the spin pattern belonging to the modes can be obtained from 
$$
\Psi_\mathbf{k} = R^{-1}(\mathbf{k}) U(\mathbf{k}) (\hat \phi L(\mathbf{k}) )^{1/2}
$$
Here $L$ is a diagonal matrix obtained as $L(\mathbf{k}) = U^{\dagger}(\mathbf{k}) X(\mathbf{k}) U(\mathbf{k})$ with the unitary matrix $U$. 
Hereby, the columns and row of $U$ should be arranged in such a way that the sign of the eigenvalues alternates, in the same way as the diagonal elements of $\hat \phi$ alternate in sign.

\paragraph{Neel antiferromagnet}
In this case, that applies to MnPS$_3$, the calculation of the magnon modes is much simpler. 
Only two magnetic ions per unit cell are required. The unitary transformation $U(\mathbf{k})$ can be written down explicitly. 
Alternatively, one may skip the para-unitary matrix, and use a Bogoliubov transform combining the operators $\hat b$ and $\hat b^\dagger$ of the two sublattices (instead of a unitary transform), see e.g. Ref.~\onlinecite{Olsen}, appendix. 
The final expression for the magnon mode reads in this case
$$
\varepsilon(\mathbf{k}) = -S \left[ \bigl(- \tilde \Delta + 3 \tilde J_1 + (\gamma_2(\mathbf{k}) -6) J_2 + 3 J_3 \bigr) \sqrt{1 - |\tilde \gamma(\mathbf{k})|^2  } \right]
$$
with
$$
\tilde \gamma(\mathbf{k}) = \frac{ \tilde J_1 \gamma_1(\mathbf{k}) + J_3 \gamma_3(\mathbf{k})}{ 3 \tilde J_1 + (\gamma_2(\mathbf{k}) -6) J_2 + 3 J_3 + (\gamma_4(\mathbf{k}) -6)  J_4}
$$
and
$$
\gamma_j = \sum_{j=1}^{\mathrm{NN}(j)} e^{ i \mathbf{k} \cdot \mathbf{R}_j }
$$
Here $\mathrm{NN}(j)$ is the number of neighboring magnetic ions in the $j$th neighbor shell, and the $\mathbf{R}_j$ are the positions of these atoms.

%\bibliographystyle{apsrev4-2}
%\bibliography{bib}

\end{document}